\documentclass[aps,twocolumn,a4paper,showpacs,nofootinbib,superscriptaddress]{revtex4}
\usepackage[dvips]{graphicx}

\begin{document}

\title{Cooper pairing of electrons and holes in graphene bilayer:\\ Correlation effects}
\author{Yu.E. Lozovik}\email{lozovik@isan.troitsk.ru}
\affiliation{Institute for Spectroscopy, Russian Academy of Sciences, 142190 Troitsk, Moscow region, Russia}
\affiliation{Moscow Institute of Physics and Technology, 141700 Dolgoprudny, Moscow region, Russia}
\author{S.L. Ogarkov}
\affiliation{National Nuclear Research University ``MEPHI'', 115409 Moscow, Russia}
\author{A.A. Sokolik}
\affiliation{Institute for Spectroscopy, Russian Academy of Sciences, 142190 Troitsk, Moscow region, Russia}
\begin{abstract}
Cooper pairing of spatially separated electrons and holes in graphene bilayer is studied beyond the mean-field
approximation. Suppression of the screening at large distances, caused by appearance of the gap, is considered
self-consistently. A mutual positive feedback between appearance of the gap and enlargement of the interaction leads to
a sharp transition to correlated state with greatly increased gap above some critical value of the coupling strength.
At coupling strength below the critical, this correlation effect increases the gap approximately by a factor of two.
The maximal coupling strength achievable in experiments is close to the critical value. This indicated importance of
correlation effects in closely-spaced graphene bilayers at weak substrate dielectric screening. Another effect beyond
mean-field approximation considered is an influence of vertex corrections on the pairing, which is shown to be very
weak.
\end{abstract}


\maketitle

\section{Introduction}
Cooper pairing of spatially separated electrons and holes due to their Coulomb attraction was proposed initially as a
possible origin of superfluidity \cite{LozovikYudson1}, Josephson effects
\cite{LozovikYudson1,Shevchenko,LozovikPoushnov} and anomalous electromagnetic phenomena
\cite{LozovikOvchinnikov,Balatsky} in coupled semiconductor quantum wells. Strong Coulomb electron-hole attraction was
supposed to maintain high critical temperature of the pairing, while spatial separation of paired electrons and holes
could prevent them from interlayer tunneling leading to their recombination and to a condensate phase fixation.
Experimental evidences of superfluid transition in two-layer semiconductor structures in strong magnetic field (pairing
of composite fermions) \cite{Eisenstein}, and without magnetic field
\cite{Snoke,Timofeev,Butov,Littlewood,Croxall,Seamons} were found several decades later.

Experimental fabrication of graphene, an atomically-thin two-dimensional form of carbon \cite{Novoselov,CastroNeto},
opens a possibility to realize electron-hole pair condensation in a spatially separated graphene bilayer. Structures
consisting of two independently gated graphene layers with common contacts and very small ($0.6\,\mbox{nm}$)
separation, have been fabricated and studied experimentally \cite{Schmidt1,Schmidt2}. A system of two independently
contacted graphene layers separated by $5\,\mbox{nm}$-thick $\mathrm{SiO}_2$ barrier have been also made \cite{Kim}.
Coulomb drag have been already studied in such system. The most promising for realization of the pairing could be
heterostructures consisting of two graphene layers separated by atomic-thin boron nitride layer. Considerable progress
in fabrication of such structures with ultra-high mobility encapsulated graphene samples have been achieved recently
\cite{Dean1,Dean2,Xue,Mayorov,Abanin}. Measurements of Coulomb drag in these structures have been reported (see
\cite{Katsnelson} and references therein).

Theoretical studies of electron-hole Cooper pairing in spatially separated graphene bilayer were presented in
\cite{Lozovik_Sokolik_JETPL,Min,Zhang,Kharitonov1,Lozovik_Sokolik_PLA,Lozovik_Sokolik_EPJB,
Bistritzer,Lozovik_Sokolik_PTRS,Kharitonov2,Mink}. It was shown that both weak and strong coupling regimes of the
pairing are achievable experimentally \cite{Lozovik_Sokolik_JETPL}. Estimates of the critical temperature
$T_\mathrm{c}$ at strong coupling
\cite{Min,Zhang,Kharitonov1,Lozovik_Sokolik_PLA,Lozovik_Sokolik_EPJB,Lozovik_Sokolik_PTRS} are very different depending
on model and approximations used: ranging from room temperature, with the unscreened Coulomb attraction taken as a
pairing potential \cite{Min,Zhang}, to unobservably small values given by Bardeen-Cooper-Shrieffer (BCS) theory with
screened interaction \cite{Kharitonov1}.

It was argued in \cite{Lozovik_Sokolik_PLA,Lozovik_Sokolik_EPJB,Mink} that the pairing in graphene bilayer at strong
coupling is multi-band, i.e. involving both conduction and valence bands of electrons and holes. The reason is that
electrons and holes in graphene are described by effective two-dimensional Dirac-type equation for massless particles
\cite{Novoselov,CastroNeto}, and thus a gap between conduction and valence band is absent. It was shown
\cite{Lozovik_Sokolik_PLA,Lozovik_Sokolik_EPJB,Mink} that $T_\mathrm{c}$ in this case can be much larger than given by
one-band BCS-like model \cite{Kharitonov1}. The study of the problem with taking into account frequency-dependence
effects was presented in \cite{Lozovik_Sokolik_PTRS}.

Mean-field approximation, used in most of these theoretical works
\cite{Lozovik_Sokolik_JETPL,Min,Zhang,Kharitonov1,Lozovik_Sokolik_PLA,Lozovik_Sokolik_EPJB,Lozovik_Sokolik_PTRS,Mink},
is known to be well-applicable at both weak- and strong-coupling sides of BCS-BEC crossover, taking place in
conventional pairing systems \cite{Nozieres,Pieri}. However, little is known about its applicability in the regime of
multi-band Cooper pairing, which takes place in graphene bilayer at strong coupling instead of the BEC (Bose-Einstein
condensate) side of the crossover \cite{Lozovik_Sokolik_EPJB,Lozovik_Sokolik_JPCS} (the BEC regime restores in graphene
bilayer if a gap is opened in its spectrum \cite{Berman,Ziegler}).

One of correlation effects arising beyond the mean-field approximation is suppression of the screening (and consequent
enlargement of electron-hole interaction) due to appearance of the gap and order parameter in the system. As supposed
in \cite{Bistritzer}, mutual positive feedback between onset of the gap and suppression of the screening can result in
appearance of two solutions of the gap equation: small gap at strong screening and large gap at weak screening.

Change of polarizability of electron system caused by appearance of a gap in its spectrum have been extensively studied
in a context of collective modes in superconductors (see, e.g., \cite{Anderson,Prange}) and phonon self-energies,
acquiring sharp features at frequencies twice larger than the gap (see \cite{Marsiglio,Kee2} and references therein).

In systems with electron-hole pairing (proposed initially as an origin of excitonic insulator \cite{Keldysh})
suppression of a system polarizability due to the pairing can change significantly the pairing interaction itself. The
question of self-consistent treatment of pairing and screening in electron-hole liquids in semiconductors have been
addressed in several works \cite{Kozlov_Maksimov,Zimmermann,Silin,Nozieres_Comte,Lozovik_Berman}.

For electron-hole graphene bilayer, the self-consistent suppression of the screening, suggested in \cite{Bistritzer},
was considered in \cite{Kharitonov2} with using the one-band approximation, however, without details of calculations
presented. The result was that this effect is negligible for a considered range of parameters and the gap equation has
always only one solution, corresponding to a very small gap. In our article, we study this effect within the model of
multi-band pairing at strong coupling. We show that (in contrast to \cite{Kharitonov2}) this effect can drastically
change characteristics of the pairing at strong enough coupling, leading to formation of very large gap.

The self-consistent suppression of the screening is described by a series of Feynman diagrams, involving the screening
of electron-hole interaction by virtual Bogolyubov excitations. Considerable part of remaining diagrams beyond
mean-field Gor'kov equations can be absorbed into renormalization of a Coulomb interaction vertex. In the theory of
superconductivity this renormalization is negligible at weak coupling due to Midgal theorem, but can be appreciable at
strong coupling, generally increasing a critical temperature \cite{Grimaldi}. Corrections to Coulomb interaction vertex
in graphene were considered in \cite{Gonzalez,Kotov,Sabio_Nilsson,Juricic}. It was shown that the vertex corrections
are large in undoped graphene \cite{Kotov}, but rather small at finite doping \cite{Sabio_Nilsson}. We calculate
numerically the simplest vertex correction and show that it is small, though acts to increase a coupling strength.

The article is organized as follows. In Sec.~2 we introduce briefly the multi-band model of the pairing, in Sec.~3 we
consider self-consistent suppression of the screening and its effect on the pairing at zero temperature. Sec.~4
presents calculations of the vertex corrections and Sec.~5 concludes the article.

\section{Multi-band description of the pairing}

Multi-band pairing of electrons and holes in graphene bilayer at strong coupling is described in detail elsewhere
\cite{Lozovik_Sokolik_PLA,Lozovik_Sokolik_EPJB,Mink}. Here we present only the formulas needed for further
calculations.

To describe the pairing, we introduce the Matsubara Green functions
$G^{(ij)}_{\gamma_1\gamma_2}(\mathbf{p},\tau)=-\langle
Tc^{(i)}_{\mathbf{p}\gamma_1}(\tau)c^{(j)+}_{\mathbf{p}\gamma_2}(0)\rangle$, where $c^{(1)}_{\mathbf{p}\gamma}\equiv
a^{(1)}_{\mathbf{p}\gamma}$, $c^{(2)}_{\mathbf{p}\gamma}\equiv a^{(2)}_{\mathbf{p},-\gamma}$,
$a^{(1)}_{\mathbf{p}\gamma}$ and $a^{(2)}_{\mathbf{p}\gamma}$ are destruction operators of electrons in electron- and
hole-doped layers respectively with the momentum $\mathbf{p}$ from conduction ($\gamma=+1$) or valence ($\gamma=-1$)
band. The bare Green functions for electrons and holes are: $G^{(ii)}_{\gamma_1\gamma_2}(\mathbf{p},i\varepsilon_n)=
\delta_{\gamma_1\gamma_2}[i\varepsilon_n-\xi^{(i)}_{\mathbf{p}\gamma_1}]^{-1}$, where
$\xi^{(1)}_{\mathbf{p}\gamma}=-\xi^{(2)}_{\mathbf{p}\gamma}\equiv\xi_{p\gamma}=\gamma v_\mathrm{F}p-\mu$ are energies
of electrons and holes measured from the chemical potentials $\mu$ and $-\mu$ in electron- and hole-doped graphene
layers respectively; $v_\mathrm{F}\approx10^6\,\mbox{m/s}$ is the Fermi velocity in graphene.

In the simplest case of $s$-wave band-diagonal pairing, all the Green functions $G^{(ij)}_{\gamma_1\gamma_2}$ are
nonzero only at $\gamma_1=\gamma_2$. Solution of the Gor'kov equations in the Cooper channel in this case is:
\begin{eqnarray}
G^{(11)}_{\gamma\gamma}(\mathbf{p},i\varepsilon_n)=\frac{u^2_{p\gamma}}{i\varepsilon_n-E_{p\gamma}}+
\frac{v^2_{p\gamma}}{i\varepsilon_n+E_{p\gamma}},\nonumber\\
G^{(21)}_{\gamma\gamma}(\mathbf{p},i\varepsilon_n)=
\frac{u_{p\gamma}v_{p\gamma}}{i\varepsilon_n-E_{p\gamma}}-
\frac{u_{p\gamma}v_{p\gamma}}{i\varepsilon_n+E_{p\gamma}},\label{G_ij}
\end{eqnarray}
and $G^{(22)}_{\gamma\gamma}(\mathbf{p},i\varepsilon_n)=-G^{(11)}_{\gamma\gamma}(\mathbf{p},-i\varepsilon_n)$,
$G^{(12)}_{\gamma\gamma}(\mathbf{p},i\varepsilon_n)=G^{(21)}_{\gamma\gamma}(\mathbf{p},i\varepsilon_n)$. Conventional
notations for coherence factors and Bogolubov energies are used:
\begin{eqnarray}
u^2_{p\gamma}=\frac12\left(1+\frac{\xi_{p\gamma}}{E_{p\gamma}}\right),\quad
v^2_{p\gamma}=\frac12\left(1-\frac{\xi_{p\gamma}}{E_{p\gamma}}\right),\nonumber\\
u_{p\gamma}v_{p\gamma}=\frac{\Delta_{p\gamma}}{2E_{p\gamma}},\quad E_{p\gamma}=
\sqrt{\xi^2_{p\gamma}+\Delta^2_{p\gamma}}.\label{coh}
\end{eqnarray}
Note that in the multi-band regime we have two kinds of the gap functions $\Delta_{p\pm}$ and Bogolyubov excitation
energies $E_{p\pm}$ corresponding to conduction and valence bands.

The gap functions are determined by equations:
\begin{eqnarray}
\Delta_{p\gamma}=-T\sum_{\gamma'i\varepsilon_n}\int\frac{d\mathbf{p}'}{(2\pi)^2}
F_{\mathbf{p}\mathbf{p}'}^{\gamma\gamma'} V(|\mathbf{p}-\mathbf{p}'|)\nonumber\\
\times G^{(21)}_{\gamma'\gamma'}(\mathbf{p}',i\varepsilon_n).\label{Sc1}
\end{eqnarray}
The factor
$F_{\mathbf{p}\mathbf{p}'}^{\gamma\gamma'}\equiv|\langle\mathbf{p}\gamma|\mathbf{p}'\gamma'\rangle|^2=
(1+\gamma\gamma'\mathbf{p}\mathbf{p}'/pp')/2$
arises in result of summation over spinor components of the effective electron wave function in graphene
\cite{Lozovik_Sokolik_EPJB}. We use here the static approximation, in which the gap $\Delta_{p\gamma}$ and the
statically screened interaction $V(q)$ are real and independent of frequency. Substituting (\ref{G_ij})--(\ref{coh})
into (\ref{Sc1}), performing summation over $\varepsilon_n=\pi T(2n+1)$ and integration over an angle of $\mathbf{p}'$,
we get at $T\rightarrow0$:
\begin{eqnarray}
\Delta_{p\gamma}=\sum_{\gamma'}\int\limits_0^\infty\frac{p\:dp}{2\pi}\,U^{(0)}_{\gamma\gamma'}(p,p')
\frac{\Delta_{p'\gamma'}}{2E_{p'\gamma'}},\label{Sc2}
\end{eqnarray}
where $U^{(0)}_{\gamma\gamma'}(p,p')=[V_0(p,p')+\gamma\gamma'V_1(p,p')]/2$ is a half-sum of half-difference of $s$- and
$p$-wave harmonics of the potential:
\begin{eqnarray}
V_l(p,p')=\int\limits_0^{2\pi}\frac{d\varphi}{2\pi}\cos(l\varphi)V\left(\sqrt{p^2+p'^2-2pp'\cos\varphi}\right).\label{V1}
\end{eqnarray}

As shown in \cite{Lozovik_Sokolik_EPJB}, the first harmonic $V_1$ can be neglected at strong coupling. In this case the
gap functions in conduction and valence bands are equal. We will find an approximate solution of (\ref{Sc2}) in a
manner similar to that used in \cite{Lozovik_Sokolik_PTRS}: we assume that $\Delta_{p\gamma}=\Delta\times f(p)$, where
$f(p)$ is some trial function, satisfying the conditions $f(p_\mathrm{F})=1$, $f(p)\propto1/p$ at $p\rightarrow\infty$.
Fixing in (\ref{Sc2}) $p=p_\mathrm{F}$ and neglecting $V_1$, we get the following algebraic equation for the gap
$\Delta$ at $T=0$:
\begin{eqnarray}
1=\int\limits_0^\infty\frac{p'\:dp'}{8\pi}V_0(p_\mathrm{F},p')f(p')
\left\{\frac1{E_{p'+}}+\frac1{E_{p'-}}\right\}.\label{Sc3}
\end{eqnarray}

\section{Self-consistent suppression of the screening}

The statically screened potential of electron-hole interaction, entering (\ref{V1}), in random phase approximation
(well-applicable in graphene bilayer due to large degree of electron state degeneracy
\cite{Kharitonov1,Kharitonov2,Apenko}) is \cite{Lozovik_Sokolik_JETPL,Mink}:
\begin{eqnarray}
V(q)=\frac{v_qe^{-qD}}{1-2v_q\Pi_0(q)+v^2_q\Pi_0^2(q)(1-e^{-2qD})},\label{V2}
\end{eqnarray}
where $v_q=2\pi e^2/\varepsilon q$ is the bare Coulomb attraction, screened by surrounding medium with dielectric
permittivity $\varepsilon$, $\Pi_0(q)$ is a static polarizability of a single graphene layer, $D$ is the interlayer
distance.

The most favorable conditions for the pairing are achieved at small interlayer separation, when $p_\mathrm{F}D\ll1$. In
this case a behavior of the system is determined only by the dimensionless parameter $r_\mathrm{s}=e^2/\varepsilon
v_\mathrm{F}\approx2.19/\varepsilon$, which specifies the coupling strength, while (\ref{V2}) reduces to
\begin{eqnarray}
V(q)=\frac{v_q}{1-2v_q\Pi_0(q)}.\label{V3}
\end{eqnarray}
The static polarization operator of doped graphene, calculated without taking into account electron-hole pairing, is
\cite{Wunsch,Hwang}
$\Pi_0(q)=g\mathcal{N}\left\{-1+\Theta(q-2p_\mathrm{F})(q/4p_\mathrm{F})G_<(2p_\mathrm{F}/q)\right\}$. Here $g=4$ is
the degeneracy factor, $\mathcal{N}=\mu/2\pi v_\mathrm{F}^2$ is the density of states (per spin projection and valley)
at the Fermi level, $G_<(x)=x\sqrt{1-x^2}-\arccos x$. The long-wavelength asymptotics $\Pi_0(q)\approx-g\mathcal{N}$
provides metallic-like screening at long distances.

Formation of a condensate of interlayer electron-hole pairs leads to appearance of a direct response of charge density
in one layer on electric field in the other layer, described by anomalous polarizability $\Pi_\mathrm{a}$. Normal,
intralayer polarizabilities $\Pi_\mathrm{n}$ of each graphene layer also change with respect to intrinsic
polarizabilities $\Pi_0$ due to appearance of the gap in the energy spectrum. The expression for $V(q)$ at
$p_\mathrm{F}D\ll1$ takes the same form as (\ref{V3}), but with the replacement
$\Pi_0(q)\rightarrow\Pi(q)\equiv\Pi_\mathrm{n}(q)+\Pi_\mathrm{a}(q)$.

\begin{figure}[t]
\begin{center}
\resizebox{0.9\columnwidth}{!}{\includegraphics{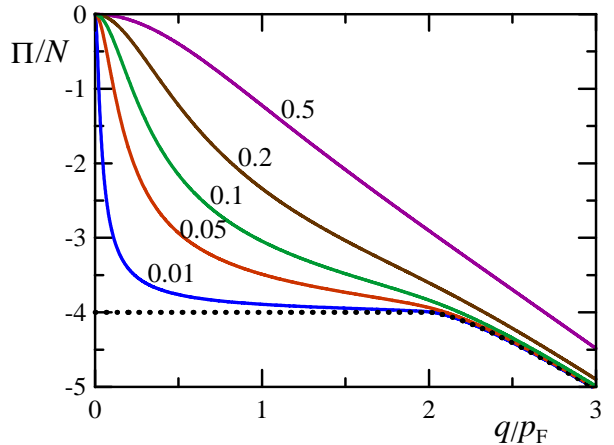}}
\end{center}
\caption{\label{Fig1}(Color online) The effective static polarizability $\Pi$ (in units of $\mathcal{N}$) in graphene
bilayer with Cooper pairing as a function of momentum $q$ at different values of $\Delta/\mu$, indicated above
corresponding curves (solid lines). Dotted line: the intrinsic static polarizability $\Pi_0$ at $\Delta=0$.}
\end{figure}

In the random phase approximation the normal $\Pi_\mathrm{n}$ and anomalous $\Pi_\mathrm{a}$ polarizabilities are
calculated as loops consisting of two normal or anomalous Green functions respectively:
\begin{eqnarray}
\Pi_\mathrm{n}(q,i\omega_n)=gT\sum_{\gamma\gamma'\varepsilon_k}\int\frac{d\mathbf{p}}{(2\pi)^2}
F_{\mathbf{p}\mathbf{p}'}^{\gamma\gamma'}G^{(11)}_{\gamma\gamma}(\mathbf{p},i\varepsilon_k)
\nonumber\\ \times G^{(11)}_{\gamma'\gamma'}(\mathbf{p}',i\varepsilon_k+i\omega_n),\label{Pn1}\\
\Pi_\mathrm{a}(q,i\omega_n)=gT\sum_{\gamma\gamma'\varepsilon_k}\int\frac{d\mathbf{p}}{(2\pi)^2}
F_{\mathbf{p}\mathbf{p}'}^{\gamma\gamma'}G^{(12)}_{\gamma\gamma}(\mathbf{p},i\varepsilon_k) \nonumber\\
\times G^{(21)}_{\gamma'\gamma'}(\mathbf{p}',i\varepsilon_k+i\omega_n),\label{Pa1}
\end{eqnarray}
where $\mathbf{p}'=\mathbf{p}+\mathbf{q}$. Substituting (\ref{G_ij}) into (\ref{Pn1})--(\ref{Pa1}), performing
frequency summations and taking a limit $T\rightarrow0$, we obtain the following expressions for the static
polarizabilities $\Pi_{\mathrm{n},\mathrm{a}}(q)\equiv\Pi_{\mathrm{n},\mathrm{a}}(q,i\omega_n\rightarrow0)$:
\begin{eqnarray}
\Pi_\mathrm{n}(q)=-g\sum_{\gamma\gamma'}\int\frac{d\mathbf{p}}{(2\pi)^2} F_{\mathbf{p}\mathbf{p}'}^{\gamma\gamma'}
\frac{u^2_{p\gamma}v^2_{p'\gamma'}+v^2_{p\gamma}u^2_{p'\gamma'}}{E_{p\gamma}+E_{p'\gamma'}},\label{Pn2}\\
\Pi_\mathrm{a}(q)=g\sum_{\gamma\gamma'}\int\frac{d\mathbf{p}}{(2\pi)^2} F_{\mathbf{p}\mathbf{p}'}^{\gamma\gamma'}
\frac{2u_{p\gamma}v_{p\gamma}u_{p'\gamma'}v_{p'\gamma'}}{E_{p\gamma}+E_{p'\gamma'}}.\label{Pa2}
\end{eqnarray}

The sum $\Pi=\Pi_\mathrm{n}+\Pi_\mathrm{a}$, playing the role of effective polarizability, is plotted in
Fig.~\ref{Fig1} as calculated numerically according to (\ref{Pn2})--(\ref{Pa2}) at different values of $\Delta$. It is
seen that polarizability of the system with Cooper pairing is suppressed at small momenta due to appearance of the gap.
A magnitude and momentum region of this suppression grow at increasing $\Delta$. The long-wavelength asymptotics
$\Pi(q)\approx-g\mathcal{N}q^2/12p_\mathrm{F}^2\Delta^2$ indicates that the screening at long distances is absent when
$\Delta\neq0$ and, moreover, transition from $\Pi(q)$ to $\Pi_0(q)$ at $\Delta\rightarrow0$ is not uniformly
continuous.

One of the consequences of this discontinuity is that the usual BCS-like recipe, involving replacement
$V_0(p,p')\rightarrow V_0(p_\mathrm{F},p_\mathrm{F})$ in the gap equation (\ref{Sc2}) becomes inapplicable, since the
Fermi-surface value $V_0(p_\mathrm{F},p_\mathrm{F})$ diverges due to absence of a long-range screening. However, the
method (\ref{Sc3}) of approximated solving of the gap equation is applicable even in this case, since the singularity
of $V_0(p_\mathrm{F},p')$ at $p'=p_\mathrm{F}$ is logarithmic and thus integrable.

\begin{figure}[t]
\begin{center}
\resizebox{0.95\columnwidth}{!}{\includegraphics{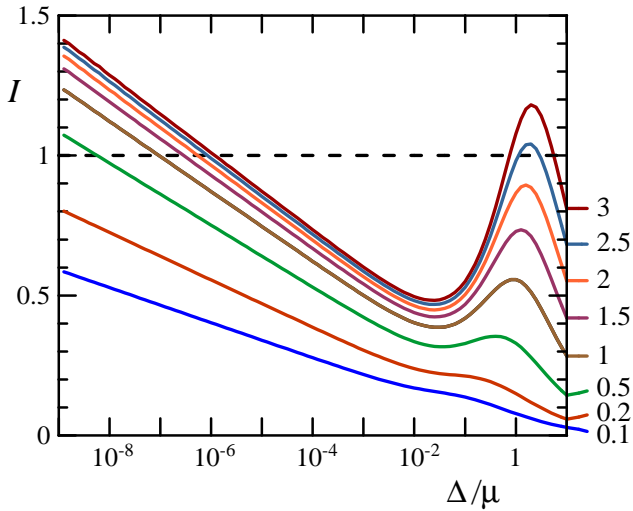}}
\end{center}
\caption{\label{Fig2}(Color online) Solid lines: the right-hand side $I$ of the gap equation (\ref{Sc3}) $I(\Delta)=1$
as a function of $\Delta$ (in logarithmic scale) at different values of $r_\mathrm{s}$, indicated in the right.}
\end{figure}

In numerical calculations, we use the trial function $f(p)=1/(|p/p_\mathrm{F}-1|+1)$ and the following approximation
for the effective polarizability: $\Pi(q)\approx-g\mathcal{N}q^2/[(12\Delta^2/v_\mathrm{F}^2)^{2/3}+q^{4/3}]^{3/2}$,
which is close to the numerically calculated $\Pi(q)$ at any $\Delta$ and retains its major features --- the correct
asymptotic at smallest $q$ and tenency to $-g\mathcal{N}$ at larger $q$.

The right-hand side $I$ of Eq.~(\ref{Sc3}) is plotted as function of $\Delta$ at various $r_\mathrm{s}$ in
Fig.~\ref{Fig2}. The points of intersection $I(\Delta)=1$ give solutions $\Delta$ of the gap equation. In the
mean-field approximation, the dependence $I(\Delta)$ should be monotonously decreasing. Correlation effects make
$I(\Delta)$ non-monotonous at $r_\mathrm{s}\gtrsim0.2$. This can potentially result in appearance of three solutions of
the gap equation. Among these solutions, only the largest $\Delta$, corresponding to the lowest ground-state energy,
will be established in the system.

\begin{figure}[t]
\begin{center}
\resizebox{0.95\columnwidth}{!}{\includegraphics{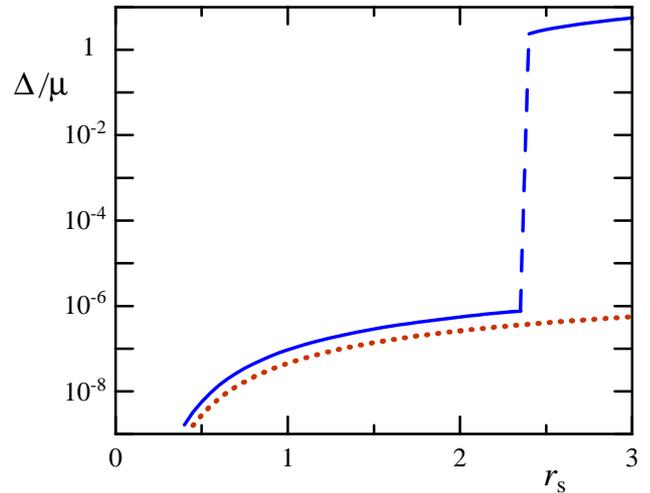}}
\end{center}
\caption{\label{Fig3}(Color online) Solid line: the largest value of the gap $\Delta$ (in logarithmic scale), found
from Eq.~(\ref{Sc3}) at different values of $r_\mathrm{s}$. At $r_\mathrm{s}\approx2.35$ the gap jumps (dashed line) to
much larger value due to appearance of three solutions of Eq.~(\ref{Sc3}). Dotted line: the gap, calculated in the
mean-field approximation.}
\end{figure}

The maximal gap $\Delta$ is plotted in Fig.~\ref{Fig3} as a function of $r_\mathrm{s}$. At small enough $r_\mathrm{s}$,
the gap value is approximately twice larger than without taking into account correlation effects. When $r_\mathrm{s}$
exceeds some critical value (about $2.35$ in our case), three solutions of (\ref{Sc3}) appear, and the maximal gap
becomes very large and comparable to the chemical potential. The critical value of $r_\mathrm{s}$ depends on details of
the model used (in particular, on a form of the trial function $f(p)$). In our case it is rather close to the maximal
$r_\mathrm{s}\approx2.19$ achievable experimentally at $\varepsilon=1$, when graphene bilayer is suspended in vacuum.
However, correlation effects can play important role in the vicinity of transition to the strongly-correlated state
even for $r_\mathrm{s}$ smaller than the critical value.

\section{Vertex corrections}

\begin{figure}[b]
\begin{center}
\resizebox{0.7\columnwidth}{!}{\includegraphics{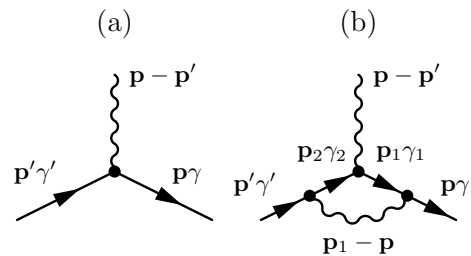}}
\end{center}
\caption{\label{Fig4}The bare vertex (a) of electron-electron Coulomb interaction and the simplest correction (b) to
it.}
\end{figure}

The bare vertex of electron-electron Coulomb interaction in graphene (Fig.~\ref{Fig4}(a)), entering the Gor'kov
equations, is $\Gamma^{(0)}_{\gamma'\gamma}(\mathbf{p}',\mathbf{p})=\langle\mathbf{p}'\gamma'|\mathbf{p}\gamma\rangle$,
where the graphene-specific angular factor is \cite{Lozovik_Sokolik_EPJB}
\begin{eqnarray}
\langle\mathbf{p}'\gamma'|\mathbf{p}\gamma\rangle=\left\{\begin{array}{rl}\cos\frac{\varphi'-\varphi}2,&\gamma=\gamma',\\
i\sin\frac{\varphi'-\varphi}2,&\gamma=-\gamma',\end{array}\right.\nonumber
\end{eqnarray}
$\varphi$ and $\varphi'$ are the azimuthal angles of $\mathbf{p}$ and $\mathbf{p}'$.

The simplest correction $\Gamma^{(1)}$ to $\Gamma^{(0)}$ is shown diagrammatically in Fig.~\ref{Fig4}(b). We take the
bare Green functions $G^{(0)}_{\gamma_2\gamma_2}(\mathbf{p}_2,i\varepsilon_2)$ and
$G^{(0)}_{\gamma_1\gamma_1}(\mathbf{p}_1,i\varepsilon_1)$ as internal electron lines of this vertex, and the Coulomb
interaction as the internal wavy line. Moreover, we consider only the static vertex, i.e. the vertex function at zero
frequencies: $\Gamma^{(1)}_{\gamma'\gamma}(\mathbf{p}',\mathbf{p})\equiv
\Gamma^{(1)}_{\gamma'\gamma}(\mathbf{p}',i\varepsilon'\rightarrow0,\mathbf{p},i\varepsilon\rightarrow0)$. After
frequency summation in the internal loop, we get in a zero-temperature limit $T\rightarrow0$:
\begin{eqnarray}
\Gamma^{(1)}_{\gamma'\gamma}(\mathbf{p}',\mathbf{p})=\sum_{\gamma_1\gamma_2}\int\frac{d\mathbf{p}_1}{(2\pi^2)}
\langle\mathbf{p}'\gamma'|\mathbf{p}_2\gamma_2\rangle\langle\mathbf{p}_2\gamma_2|\mathbf{p}_1\gamma_1\rangle\nonumber\\
\times\langle\mathbf{p}_1\gamma_1|\mathbf{p}\gamma\rangle
V(|\mathbf{p}_1-\mathbf{p}|)\frac{\Theta(\xi_{p_1\gamma_1})-
\Theta(\xi_{p_2\gamma_2})}{\xi_{p_1\gamma_1}-\xi_{p_2\gamma_2}},\label{vert1}
\end{eqnarray}
where $\mathbf{p}_2=\mathbf{p}_1+\mathbf{p}'-\mathbf{p}$ and $\Theta(x)$ is a unit step function.

The integral in (\ref{vert1}) diverges at small $q$ if we take the bare Coulomb interaction $V(q)=2\pi e^2/\varepsilon
q$. Thus we will take the statically screened interaction (\ref{V3}) as $V(q)$.

\begin{figure}[t]
\begin{center}
\resizebox{0.9\columnwidth}{!}{\includegraphics{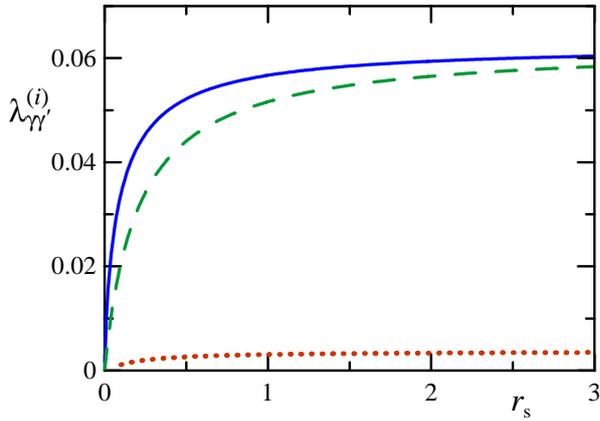}}
\end{center}
\caption{\label{Fig5}(Color online) The dimensionless second-order correction $\lambda^{(1)}_{++}$ (dotted line) to the
coupling constant as a function of $r_\mathrm{s}$ in comparison with the first-order intraband
$\lambda^{(0)}_{\gamma\gamma}$ (solid line) and interband $\lambda^{(0)}_{\gamma,-\gamma}$ (dashed line) coupling
constants.}
\end{figure}

Substitution of the vertex correction (\ref{vert1}) into Gor'kov equations results in the correction to the potential
\begin{eqnarray}
U^{(1)}_{\gamma\gamma'}(p,p')=\int\limits_0^{2\pi}\frac{d\varphi'}{2\pi}\langle\mathbf{p}\gamma|\mathbf{p'}\gamma'\rangle
\Gamma^{(1)}_{\gamma'\gamma}(\mathbf{p}',\mathbf{p}),\label{vert2}
\end{eqnarray}
added to $U^{(0)}_{\gamma\gamma'}(p,p')$ in the gap equation (\ref{Sc2}). To estimate the effect of vertex corrections
on the pairing, we compare the second-order correction to the coupling constant on the Fermi surface
$\lambda^{(1)}_{++}=\mathcal{N}U^{(1)}_{++}(p_\mathrm{F},p_\mathrm{F})$ with its first-order value
$\lambda^{(0)}_{\gamma\gamma'}=\mathcal{N}U^{(0)}_{\gamma\gamma'}(p_\mathrm{F},p_\mathrm{F})$.

In Fig.~\ref{Fig5} the conduction-band component of the correction $\lambda^{(1)}_{++}$ is plotted as function of
$r_\mathrm{s}$. Its other components ($\lambda^{(1)}_{--}$ and $\lambda^{(1)}_{\gamma,-\gamma}$) are very close to it
and thus are not shown. For comparison, the first-order coupling constants in the intra-band
($\lambda^{(0)}_{\gamma\gamma}$) and inter-band ($\lambda^{(0)}_{\gamma,-\gamma}$) channels are also plotted (these
quantities were studied in detail in \cite{Lozovik_Sokolik_EPJB}). It is seen that the second-order vertex corrections
amount to about $5\%$ of the first-order coupling constants and thus can be neglected.

\section{Conclusions}
We have considered the correlation effects in Cooper pairing of spatially separated electrons and holes in graphene
bilayer at strong coupling. The first effect considered is the self-consistent suppression of the screening of
electron-hole interaction due to appearance of the gap and order parameter. Its most remarkable consequence is an
absence of screening at long distances at any nonzero gap, that makes a usual BCS method inapplicable due to divergence
of the interaction at the Fermi surface.

Thus we performed numerical calculations with full momentum integration to solve the gap equation. We found that, at
small enough coupling strengths, correlations increase the gap by a factor of two. However, at coupling strength above
some threshold the gap sharply increases by several orders of magnitude, which indicates transition of the system to a
strongly-correlated state. The critical value of the coupling strength depends on details of the theoretical model. In
our case it turned out to be only slightly larger than the maximal value achievable in experiments.

We have also considered the role of vertex corrections at strong coupling. The simplest correction to the interaction
vertex was calculated numerically in the static approximation and was shown to enhance the pairing, but only on about
5\%.

The work was supported by Russian Foundation for Basic Research. One of the authors (A.A.S.) was also supported by the
Dynasty Foundation and by the grant of the President of Russian Federation for Young Scientists MK-5288.2011.2.

\end{document}